This paper is edited in the standard Latex form.

\documentstyle[11pt]{article}

\evensidemargin=\oddsidemargin
\begin{document}

\begin{flushright}
{\large \sf  TUIMP-TH-94/56\\
             VPI-IHEP-93-15\\
             December, 1993 }
\end{flushright}
\bigskip
\bigskip
\begin{center}
{\Large \bf PROOF OF THE EQUIVALENCE THEOREM
            IN THE CHIRAL LAGRANGIAN FORMALISM }
\footnote{ Work supported by the National Natural Science Foundation of China
and the U.S. Department of Energy under grant DEFG0592ER40749.}
\\[1.5cm]
{\bf Hong-Jian He} \\[0.3cm]
Department of Physics and Institute of High Energy Physics,\\
Virginia Polytechnic Institute and State University, \\
Blacksburg, Virginia 24061-0435, U.S.A.\footnote{ Mailing address.}\\ [0.5cm]
{\bf Yu-Ping Kuang} \\[0.3cm]
China Center of Advanced Science and Technology(World Laboratory), P.O.Box
8730, Beijing 100080, China; and Institute of Modern Physics, Tsinghua
University, Beijing 100084, China$^2$; and Institute of Theoretical Physics,
Academia Sinica, Beijing  100080, China \\[0.5cm]
{\bf Xiaoyuan Li}\\[0.3cm]
Institute of Theoretical Physics, Academia Sinica, Beijing 100080, China$^2$
\end{center}

\newpage
\null
\centerline{\bf Abstract}
\bigskip
\begin{sf}

A general proof of the equivalence theorem in electroweak theories with the
symmetry breaking sector described by the chiral Lagrangian is given in the
$R_{\xi}$ gauge by means of the Ward-Takahashi identities. The precise form of
the theorem contains a modification factor $C_{mod}$ associated with each
external Goldstone boson similar to that in the standard model. $C_{mod}$ is
exactly unity in our previously proposed renormalization scheme,
{\it Scheme-II}.

\vspace{3.0cm}
\end{sf}

\newpage
\begin{sf}

The equivalence theorem (ET) is very useful in relating the electroweak
symmetry breaking mechanism to the longitudinal weak-boson scattering
experiments and in simplifying the calculations of multiple longitudinal
weak-boson amplitudes$^{[1]}$.
So far, the form of the ET has only been carefully
studied within the framework of the
standard model (SM)$^{[2]-[5]}$. Since the
electroweak symmetry breaking mechanism is still unclear, the study should
also include the probe of mechanisms beyond the SM. This needs the knowledge
of the ET in the corresponding theory, which has not been rigorously proved
yet. Regardless of the details of the symmetry breaking mechanism, the
dynamics of the would-be Goldstone bosons (GB)
can be effectively described by
a local chiral Lagrangian to certain order in the momentum  expansion with
unknown coefficients. In this letter, we present briefly the general proof of
the ET in theories whose GB dynamics is described by such a chiral Lagrangian.
The proof is given in the general $R_{\xi}$ gauge by means of the
Ward-Takahashi (WT) identities obtained
from the BRST invariance of the theory,
which is essentially parallel to that given in Ref.[5] for the SM. A longer
paper$^{[6]}$ following this will present the details.

For simplicity, we neglect the Weinberg angle and consider the $SU(2)_L$ gauge
theory of weak interactions (The generalization to the complete
$SU(2)\times U(1)$ electroweak theory is straightforward as is shown in
Ref.[5] for the SM.). Moreover, at the moment, we concentrate our attention
only upon the bosonic sector which is essential in the proof. Let $W^a_{\mu}$
be the $SU(2)_L$ gauge boson, $\pi^a$ be the GB whose dynamics is described by
the $SU(2)_L\times SU(2)_R$  chiral Lagrangian, and $c^a$ and $\bar c^a$ be
the ghost and antighost fields, respectively. The chiral Lagrangian can be
formulated via the nonlinear realization, say
\begin{eqnarray}                       
U(\pi^a)\,=\,\exp [i\tau ^a\pi ^a/f_{\pi}] ~,
\end{eqnarray}
where  $f_{\pi}$  is the GB decay constant which is equal to the vacuum
expectation value (VEV) breaking the gauge symmtry, and $\tau^a/2$  is the
generator of $SU(2)$. Define
\begin{equation}          
\begin{array}{ll}
W_{\mu}\,\equiv\,-iW^a_{\mu}\tau^a/2 ~,\;\;\;\;\;\;\;\;
W_{\mu\nu}\,\equiv\,\partial_{\mu}W_{\nu}\,-\,\partial_{\nu}W_{\mu}\,+
\,g[W_{\mu},W_{\nu}] ~,\\
D_{\mu}\,\equiv\,\partial_{\mu}\,+\,gW_{\mu} ~,\;\;\;\;\;\;\;\;
{\cal D}_{\mu}\,\equiv\,\partial_{\mu}\,+\,g[W_{\mu},\;] ~,\\
V_{\mu}\,\,\equiv\,(D_{\mu}U)U^\dagger\,
=\,gW_{\mu}\,+\,(\partial_{\mu}U)U^\dagger ~.
\end{array}
\end{equation}
The Lagrangian for the bosonic sector can then be written as
\begin{eqnarray}                
{\cal L}_{eff}\,=\,{\cal L}\,+\,{\cal L}_{gf}\,+\,{\cal L}_{FP} ~,
\end{eqnarray}
\noindent
where ${\cal L}_{gf}$ and ${\cal L}_{FP}$ are the gauge fixing
and the Faddeev-Popov terms,
and  ${\cal L}\,=\,{\cal L}_W\,+\,{\cal L}_{GB}$  is the
Lagrangian for the gauge and GB fields, with$^{[7]}$
\begin{equation}                 
\begin{array}{ll}
{\cal L}_W\,=\,{\displaystyle\frac 12}tr(W^{\mu\nu}W_{\mu\nu}) ~,\\
{\cal L}_{GB}\,=\,\sum_n{\cal L}^{(2n)}_s\,=\,{\cal L}^{(2)}_s\,+
\,{\cal L}^{(4)}_s\,+\cdots\cdots,\\
{\cal L}^{(2)}_s\,=\,\displaystyle\frac{f^2_{\pi}}{4}
tr[(D_{\mu}U)(D^{\mu}U)^\dagger ]~,\\
{\cal L}^{(4)}_s\,=\,\alpha_1tr(V^{\mu}V_{\mu})tr(V^{\nu}V_{\nu})\,+
\alpha_2tr(V_{\mu}V_{\nu})tr(V^{\mu}V^{\nu})\,+
\,\alpha_3gtr(W_{\mu\nu}[V^{\mu},V^{\nu}]\,+
\,\alpha_4tr[({\cal D}_{\mu}V^{\mu})({\cal D}_{\nu}V^{\nu})],\\
\cdots\cdots,\\
\end{array}
\end{equation}
\noindent
in which  $\alpha_1,\cdots,\alpha_4$ are unknown coefficients. For the general
$R_{\xi}$ gauge$^{[5]}$,
\begin{equation}                 
\begin{array}{ll}
{\cal L}_{gf} & = \displaystyle\frac{1}{2\xi}F^2_a~,\;\;\;\;\;\;\;\;\;\;
F_a\,=\partial_{\mu}W_a^{\mu}\,+\,\xi\kappa\pi_a~,\\
{\cal L}_{FP} & =-\int d^4y\bar c_a(x)$$\displaystyle\frac{\delta F_a(x)}
{\delta\theta_b(y)}c_b(y) ~,
\end{array}
\end{equation}
where $\theta_b$ is the infinitesimal gauge transformation parameter.

In the path integral formalism,
\begin{equation}                  
Z\,=\,\int {\cal D}W\,{\cal D}U\,{\cal D}c\,{\cal D}{\bar c}\,\,
\exp\,i\int d^4x\,{\cal L}_{eff}~
\,=\,\int {\cal D}W\,{\cal D}\pi\,{\cal D}c\,{\cal D}
{\bar c}\,{\Delta}_J(\pi)\,\,\exp\,i\int d^4x\,{\cal L}_{eff} ~,
\end{equation}
\noindent
where ${\Delta}_J(\pi)$ is the Jacobian for the change of variable
$~U\rightarrow\pi$ and can be ignored in dimensional
regularization$^{[8]}$ since it can be written as
$~{\Delta}_J(\pi)=\exp\,i\,\int d^4x\,
[-i{\delta^4(0)}\ln{\Delta}_J(\pi)]~$
in which the $\delta$-function vanishes in dimensional regularization.
To consider the BRST transformation,
it is convenient to introduce an auxiliary
field $B_a$ by inserting a Gaussian type integral into (6), i.e.
\begin{equation}               
\begin{array}{ll}
Z\,=\,\int {\cal D}W\,{\cal D}\pi\,{\cal D}c\,{\cal D}{\bar c}\,{\cal D}B\,\,
\exp\,i\,\int d^4x\,[{\cal L}_{eff}\,+\,{\displaystyle\frac{1}{2\xi}}(B_a\,
+\,F_a)^2]\\
\;\;\;\;=\,\int {\cal D}W\,{\cal D}\pi\,{\cal D}c\,{\cal D}{\bar c}\,
{\cal D}B\,\,\exp\,i\,\int d^4x\,[{\cal L}\,+\,{\bar {\cal L}_{gf}}\,+\,
{\cal L}_{FP}] ~,\\
\end{array}
\end{equation}
\noindent
where the new effective gauge fixing term is
\begin{equation}               
\bar {\cal L}_{gf}\,\equiv {\cal L}_{gf}\,+\,{\displaystyle\frac{1}{2\xi}}
(B_a\,+\,F_a)^2
=\,{\displaystyle\frac{1}{2\xi}}B_a^2\,+\,{\displaystyle
\frac{1}{\xi}}B_aF_a~.
\end{equation}
The Euler-Lagrange equation for $B_a$ is
\begin{eqnarray}               
B_a\,=\,-F_a ~,
\end{eqnarray}
\noindent
with which ~${\bar {\cal L}_{gf}}$~
is equivalent to ~${\cal L}_{gf}$~. It
is easy to check that ${\cal L}\,+\,{\bar {\cal L}_{gf}}\,+\,{\cal
L}_{FG}$ is invariant under the BRST transformation$^{[9]}$
\begin{equation}             
\begin{array}{ll}
s(W^a_{\mu})\lambda \,=\,R^a_{{\mu}b}(W)\,c^b\,\lambda ~, &
R^a_{{\mu}b}\,\equiv -{\partial}_{\mu}{\delta}^{ab}\,+\,g{\epsilon}^{abc}
W^c_{\mu} ~,\\
s({\pi}^a)\lambda\,\,\,=\,R^a_b(\pi)\,c^b\,\lambda ~, &
R^a_b(\pi)\,\equiv {\displaystyle\frac{gf_\pi}{2}}
\left[ {\epsilon}^{abc}\b{$\pi$}^c - {\delta}^{ab}\b{$\pi$}\cot\b{$\pi$}
\,+\,(\b{$\pi$}\cot\b{$\pi$}-1)\displaystyle\frac{\b{$\pi$}^a
\b{$\pi$}^b}{\b{$\pi$}^2}\,\right]~, \\
s(c^a)\lambda\,\,\,\,
=\,-{\displaystyle\frac 12}{\epsilon}^{abc}\,c^b\,c^c\lambda~, & \\
s({\bar c}^a)\lambda\,\,\,\,=\,\xi^{-1}B^a\lambda ~, & \\
s(B^a)\,\,\,\,\,=\,0~, & \\
\end{array}
\end{equation}
where $\lambda$ is an infinitesimal Grassmann parameter, and
$~\b{$\pi$}^a\equiv\pi^a/f_{\pi}~$,~$~\b{$\pi$}\equiv\sqrt{
\b{$\pi$}^a\b{$\pi$}^a}~$.  With the symbols in (10), the
explicit formula for ${\cal L}_{FP}$ reads
\begin{eqnarray}            
{\cal L}_{FP}\,=\,-{\bar c}^a\,{\displaystyle\frac{{\delta}F^a}
{{\delta}W^b_{\mu}}}\,R^c_{{\mu}b}(W)\,c^c\,-\,{\bar c}^a\,
{\displaystyle\frac{{\delta}F^a}{{\delta}\pi^b}}\,R^c_b(\pi)\,c^c.
\end{eqnarray}

There are interaction terms with
dimension $>4$~ operators in ${\cal L}_{GB}$
and ${\cal L}_{FP}$, which are non-renormalizable in the perturbation sense.
Conventionally, the renormalization of this kind of theory is proceeded to a
certain order $p^{2n}$ in the momentum expansion, i.e. taking account of
necessary counter terms to order-$p^{2n}$ to cancel the corresponding
divergences$^{[10,9]}$. To make the renormalization procedure BRST invariant,
the counter terms should be BRST invariant. The ghost-independent counter
terms have been systematically constructed in a gauge-invariant (also
BRST-invariant) way in Ref.[7]. ( The $~p^4$-order counter terms of such kind
have been explicitly listed in the above eq.(4). )
Furthermore, some new BRST-invariant,
ghost-dependent counter terms should be added to ${\cal L}_{FP}$ in the
general $~R_\xi~$ gauge ( except $~\xi = 0~$ )
since the second term in (11) contains {\it non-renormalizable}
GB-ghosts interactions. Due to the nilpotency of the BRST
transformation operator $s$, these BRST invariant counter terms can be written
down by applying $s$ to certain field operators$^{[6]}$.
The following general proof does not
concern the explicit form of these counter terms, and we shall present
them up to order-$p^4$ in Ref.[6]. Having all these, we can follow Ref.[5] to
analyze the renormalization of the unphysical sector in the theory, which is
important in formulating the ET. Define the
renormalization constants
\begin{equation}              
\begin{array}{ll}
{\pi}^a_0\,=\,{\sqrt Z_{\pi}}\,{\pi}^a,\;\;\;\;c^a_0\,=\,Z_c\,c^a,\;\;
\;\;{\bar c}^a_0\,=\,{\bar c}^a,\\
\xi_0\,=\,Z_{\xi}\,{\xi}~,\;\;\;
\;\kappa_0\,=\,Z_{\kappa}\,{\kappa}~,\\
\end{array}
\end{equation}
\noindent
where the subscript "$0$" denotes unrenormalized quantities. They are
constrained by the following WT identities obtained from the BRST invariance
of the theory,
\begin{equation}                                      
\begin{array}{l}
ik^\mu[i{\cal D}^{-1}_{0,\mu\nu}(k)+ \xi^{-1}_0 k_\mu k_\nu]+
M_{W0}\hat{C}_0(k^2)[i{\cal D}^{-1}_{0,\pi\nu}(k)- i
\kappa_0 k_\nu]=0~~,\\
ik^\mu[-i{\cal D}^{-1}_{0,\pi\mu}(k)+i \kappa_0 k_\mu]+
M_{W0}\hat{C}_0(k^2)[i{\cal D}^{-1}_{0,\pi\pi}(k)+\xi_0 \kappa^2_0]=0~~,\\
iS^{-1}_{0, ab}(k)=[1+\Delta_3(k^2)][k^2-\xi_0\kappa_0 M_{W0}\hat{C}_0(k^2)]
\delta_{ab}~~,
\end{array}
\end{equation}
in which the  $\cal D$'s  are physical propagators, and
\begin{equation}                                     
\hat{C}_0(k^2)=\frac{1+\Delta_1(k^2)+\Delta_2(k^2)}{ 1+\Delta_3(k^2)}~,
\end{equation}
\noindent
with
\begin{equation}                                      
\begin{array}{ll}
\Delta_1(k^2)\delta^{ab} &
\equiv
F.T.<0\,|T[\b{$\pi$}_0(y){\cot\b{$\pi$}}_0(y)-1]
\left[\delta^{bd}-\displaystyle
\frac{\b{$\pi$}_{0}^b(y)\b{$\pi$}_{0}^d(y)}{\b{$\pi$}_0^2}\right]\,
c_0^d(y)|\,\bar{c}^a_0(x)>~,\\
\Delta_2(k^2)\delta^{ab} &
\equiv -\displaystyle\frac{g_0}{2M_{W0}}\epsilon^{bcd}\,
\int_q<0\,|\pi^c_0(-k-q)\,c^d_0(q)|\,\bar{c}^a_0(k)> ~,\\
ik_\mu\Delta_3(k^2)\delta^{ab} &
\equiv -g_0\epsilon^{bcd}\,\int_q<0\,|W^c_{\mu0}(-k-q)\,c^d_0(q)
|\,\bar{c}^a_0(k)> ~,
\end{array}
\end{equation}
where "F.T." denotes " Fourier transformation " and
$~\int_q\equiv\mu^{\epsilon}\int\displaystyle\frac{d^Dq}{(2\pi)^D}~$
with $~D=4-\epsilon~$ in dimensional regularization.
These $\Delta$'s come from the non-factorized parts of the VEV's containing
the BRST transformed fields, which vanish at tree level and get nonvanishing
contributions from loop diagrams. The nonvanishing  $\Delta$'s  make
$~\hat C_0(k^2)\neq 1~$  at loop level. After the above renormalization, (13)
becomes
\begin{equation}                                   
\begin{array}{l}
ik^\mu [i{\cal D}^{-1}_{\mu\nu}(k)+
\frac{Z_W}{Z_\xi}\xi^{-1}k_\mu k_\nu]+
Z_{M_W}\left(\frac{Z_W}{Z_\pi}\right)^{\frac{1}{2}}\hat{C}_0(k^2)M_W[i{\cal
D}^{-1}_{\pi\nu}(k)-  Z_\kappa Z^{\frac{1}{2}}_W Z^{\frac{1}{2}}_\pi
 ik_\nu\kappa]=0~~,\\
ik^\mu [-i{\cal D}^{-1}_{\pi\mu}(k)+
Z_\kappa Z^{\frac{1}{2}}_W
Z^{\frac{1}{2}}_\pi ik_\mu\kappa ]+
Z_{M_W}\left( \frac{Z_W}{Z_\pi}\right) ^{\frac{1}{2}}\hat{C}_0(k^2)M_W
[i{\cal D}^{-1}_{\pi\pi}(k)+Z^2_\kappa Z_\xi  Z_\pi  \xi\kappa^2 ]=0
{}~~,\\
iS^{-1}_{ab}(k)=Z_c[1+\Delta_3(k^2)][k^2-\xi\kappa M_W Z_\xi Z_\kappa
Z_{M_W}\hat{C}_0(k^2)]\delta_{ab}~~,
\end{array}
\end{equation}
where $~{\cal D}_{o,\mu\nu}=Z_W{\cal D}_{\mu\nu},~~
{\cal D}_{o,\pi\nu}=Z^{\frac{1}{2}}_\pi Z^\frac{1}{2}_W
{\cal D}_{\pi \nu}$~, etc..
We can then define the renormalized quantity$^{[5]}$
\begin{equation}                                      
\hat{C}(k^2)\equiv
\left( \frac{Z_W}{Z_\pi}\right)^{\frac{1}{2}} Z_{M_W}\hat{C}_0(k^2)~~.
\end{equation}
\noindent
The finiteness of the renormalized quantities in (16) leads to the
following constraints on the renormalization constants
\begin{equation}                                      
\begin{array}{ll}
Z_\xi = \Omega_\xi Z_W~~, & Z_\kappa=\Omega_\kappa Z^{\frac{1}{2}}_W
Z^{-\frac{1}{2}}_{\pi}Z_\xi^{-1}~~,\\
Z_\pi=\Omega_\pi Z_W Z^2_{M_W} \hat{C}_0(sub.~ point)~~, &
Z_c=\Omega_c[1+\Delta_3(sub.~ point)]^{-1}~~,
\end{array}
\end{equation}
where $~\Omega_\xi~$, $~\Omega_\kappa~$,
$~\Omega_\pi~$ and $~\Omega_c~$ are
finite constants to be determined by the subtraction conditions.
The above  expressions
 are essentially the same as those in the SM given in Ref.[5] except
that the formula for  $\Delta_1$  is different. In the two convenient
renormalization schemes, {\it Scheme-I} and {\it Scheme-II}, proposed
in Ref.[5],  $\hat C(k^2)$ is simply
\begin{equation}                                                   
\hat C(k^2)=\left\{
\begin{array}{ll}
\Omega^{-1}_\kappa & {\sf ,~~in~~Scheme-I~~with~~\kappa=M_W~~ and~~\xi=1~~;}\\
1             &{\sf ,~~in~~Scheme-II~~ with~~\kappa=\xi^{-1}M_W~~.}
\end{array}
\right.
\end{equation}
\noindent
As has been proved in Ref.[5], the on-shell value of $~\hat C(k^2)$ is
proportional to the modification factor $~C_{mod}$ appearing in the ET,
so that  {\it Scheme-II} is the most convenient
scheme in practical calculations.

The general proof of the ET given in Sec.III of Ref.[5] includes the use of
the Slavnov-Tayler (ST) identity$^{[3][5]}$
\begin{equation}                                                  
<0\,|F^{a_1}_0(k_1)\,F^{a_2}_0(k_2)\cdots F^{a_n}_0(k_n)\,\Phi|\,0>\,=\,0~,
\end{equation}
\noindent
and doing the amputation and renormalization for (20). These procedures does
not concern the explicit formula for the $~\Delta$'s  so that they can be
exactly applied to the present case without any modifications. We are not
going to repeat the procedures here and simply quote the result in Ref.[5].
The obtained precise form of the ET is$^{[5]}$
\begin{eqnarray}                                                   
T(W_L^{a_1},W_L^{a_2},\cdots,W_L^{a_n})=C_{mod}^{a_1}C_{mod}^{a_2}\cdots
C_{mod}^{a_n}T(-i\pi ^{a_1},-i\pi ^{a_2},\cdots,-i\pi ^{a_n})+O(M_W\,/E)~,
\end{eqnarray}
where $W_L^a$ is the longitudinal component of $W_{\mu}^a$, and the
modification factor $~C_{mod}$ is
\begin{equation}                                                     
C_{mod}=\frac{M_W}{M^{phys}_W} {\hat C}((M^{phys}_W)^2)~.
\end{equation}
\noindent
In (22), $~M^{phys}_W$  is the physical mass of the $W$-boson which may be
different from $M_W$ in some renormalization schemes, and  $M^{phys}_W\,
=\,M_W$ in the on-shell scheme.
Here we concentrate our attention mainly upon the important quantization
and renormalization effects in the precise formulation of ET. In the
chiral Lagrangian formalism, the amplitude is energy dependent to each
order in the momentum expansion and is valid in the region
$~E\ll 4\pi f_{\pi}~$. In practical applications, when ignoring the terms
containing $~v^{\mu}=\epsilon_L^{\mu}-\epsilon_S^{\mu}=O(M_W/E)~$ to
obtain (21), certain conditions ( e.g. $~M_W\ll E\ll 4\pi f_{\pi}~$ ) ensuring
the largest $v_{\mu}$-suppressed term
to be much smaller than the smallest term
kept in the 1st term on the RHS of (21) are required. The technical detail
of the conditions is presented in Ref.[6].
In our {\it Scheme-I} and {\it Scheme-II},
$~\hat C((M^{phys}_W)^2)$ has been given in (19), so that we have
\begin{equation}                                                    
C_{mod}=\left\{
\begin{array}{ll}
\Omega^{-1}_\kappa &
{\sf ,~~in~~ Scheme-I~~ with~~ \kappa=M_W~~ and~ ~\xi=1~;}\\
1             &{\sf,~~in ~~ Scheme-II~~ with~~ \kappa=\xi^{-1}M_W~.}
\end{array}
\right.
\end{equation}
\noindent
Therefore in {\it Scheme-II}, $~C_{mod}$ is exactly unity and the ET
takes its simplest naive form. Eqs.(21)-(23) are
{\it the main conclusions} of this paper.

Finally, we briefly discuss the contributions from the fermions. The fermion-W
coupling is the standard gauge coupling which is perturbatively renormalizable,
 so that it does not cause any complication in the renormalization. The
fermion-GB coupling is more complicated in the nonlinear chiral Lagrangian
formalism. It contains perturbatively non-renormalizable terms and thus BRST
invariant counter terms including fermion fields are needed, however, this
does not affect the validity of the above general proof which does not concern
the detailed expressions of the counter terms.
In a recent paper$^{[11]}$, Donoghue and
Tandean claimed that the ET would be violated in a kind of technicolor (TC)
type model with "global anomaly" through the triangle fermion loop (TFL)
contributions to the scattering amplitutes. They took a toy model with one
family of fermions, and the GB couples only to the quarks (corresponding to
the technifermions in the TC model) but not to the leptons (corresponding  to
the ordinary fermions in the TC model). They compared the TFL contributions to
the neutral $~Z^0_L-\gamma-\gamma^\ast~$ and $~GB-\gamma-\gamma^\ast~$
amplitudes ( where $\gamma^\ast~$ is a virtual photon ).
Their argument is that the whole family of fermions contribute to the TFL in
the $~Z^0_L-\gamma-\gamma^\ast~$ amplitude so that their sum is zero due to
the gauge anomaly cancellation, while only quarks contribute to the TFL in the
$~GB-\gamma-\gamma^\ast~$ amplitude just like the $~\pi^0\rightarrow
\gamma\gamma~$
amplitude which is nonvanishing due to the global anomaly, so that the ET is
violated. This is confusing since the ET is a general consequence of the
WT(ST) identities which is irrelevant to the details of the TFL diagrams.
There have been several authors clarifying this issue and showing that the ET
actually holds in this toy model$^{[12,13]}$. We would like to mention briefly
some key points here and show that the correct result is
{\it consistent} with our
general formula (21)-(22). The key points are:
\begin{enumerate}
\item                        
The $~\pi^0\rightarrow \gamma\gamma~$ amplitude is related to the global
anomaly through Sutherland-Veltman's theorem$^{[14]}$ in the zero momentum
limit due to its low energy nature, while the ET concerns the high energy
problems  ($E>>M_W$) so that the $~GB-\gamma-\gamma^\ast~$
amplitude is not related to the global anomaly
but is a pseudoscalar-vector-vector (P-V-V) triangle quark-loop amplitude and,
to lowest order, is proportional to the effective Yukawa coupling coupling
constant $~f_{quark}$.
\item                        
At high energies, the longitudinal polarization vector $\epsilon_L^{\mu}$ is
approximately proportional to $k^{\mu}$, so that the
$~Z^0_L-\gamma-\gamma^\ast~$
amplitude is proportional to the
divergence of the axial vector-vector-vector
(A-V-V) TFL amplitude, which contains a
{\it normal} term and an {\it anomaly} term.
The cancellation of the gauge anomaly means that the anomaly term vanishes. So
that the $~Z^0_L-\gamma-\gamma^\ast~$ amplitude
{\it equals to its normal term} which
does not vanish and is equal to the sum of $2m_{fermion}$ times the P-V-V
amplitude over all fermions. Since the leptons does not couple to the GB, they
are massless and thus do not contribute to the normal term. So that the normal
term contains only the contribution from the quarks, which is closely related
to the $~GB-\gamma-\gamma^\ast~$ amplitude.
\item                       
Let $~v~$ be the VEV breaking the symmetry. To lowest order, $~m_{quark}=
f_{quark}v$. By using this relation, it is straightforward to check explicitly
that, to lowest (1-loop) order,
the normal term of the $~Z^0_L-\gamma-\gamma^\ast~$ amplitude
just equals to the $~GB-\gamma-\gamma^\ast~$ amplitude
after ignoring $O(M_W/E)$ terms in the high energy limit.
This is the desired result for the validity of the ET.
\end{enumerate}
Here we emphasize that the modifaction factor
$C_{mod}$ for each GB external
line in ET (cf. (21)(22)) {\it generally exists} even in the presence of
fermion loops since our proof of ET is based only on the general WT(ST)
identities. When the ET is used to relate the $A_L-V-V^\ast$ and
$GB-V-V^\ast$ amplitudes,  $C_{mod}$ must be included as what is shown in
(21). In 1-loop calculations, the $~O(1-loop)~$  modifiaction $~C_{mod}-1~$
cannot be ignored if the tree level vertices $A_L-V-V^\ast$ and $GB-V-V^\ast$
do not vanish. The above example is a special case in which the tree level
$~Z^0_L-\gamma-\gamma^\ast~$ and
$~GB-\gamma-\gamma^\ast~$ vertices vanish, so
that $C_{mod}-1$ only causes $~O(2-loop)~$
modifications which are negligible in
the $~O(1-loop)~$ calculation. But if we consider amplitudes like
$~W^+_L-W^-_L-\gamma^\ast~$ and $~\pi^+-\pi^--\gamma^\ast~$, $C_{mod}$ will
certainly modify the naive ET even at 1-loop level due to the corresponding
non-vanishing tree level vertices.
So we finally conclude that the modification factor
$~C_{mod}~$ is generally different from unity and must be carefully included
in the application of the ET (cf. (21-23))
unless the {\it renormalization  Scheme-II}~ is adopted.

\noindent
{\bf Acknowledgements}  Two of us (H.J.H. and Y.P.K.) would like to thank
Prof. Howard Georgi for reading some of our previous unpublished
results$^{[12]}$ and for his kind suggestions.
H.J.H. is grateful to Profs. Lay Nam Chang, Chia-Hsiung Tze and C.P. Yuan
for helpful discussions.

\noindent
{\bf References}

\begin{enumerate}
\item                    
For examples,  W.J. Marciano, G.Valencia, and
S. Willenbrock, Phys. Rev. {\bf D40}(1990)1725; M.J.G. Veltman and
F.J. Yndurain, Nucl. Phys. {\bf B325}(1989)1; G. Passarino, Nucl. Phys.
{\bf B343}(1990)31; L. Durand, J.M. Johnson, and P.N. Maher, Phys. Rev.
{\bf D44}(1991)127; C. Duun and T.M. Yan, Nucl. Phys. {\bf B352}(1991)402;
X. Li, Report No. MPI-PAE/PTh 84/84(unpublished); A. Dobado, M.J. Herrero, and
T.N. Truong, Phys. Lett. {\bf 235B}(1990)129; {\bf 235B}(1990)135;
{\bf 237B}(1990)457;
J.F. Donoghue and C.Ramirez, Phys. Lett. {\bf 234B}(1990)361;
J. Bagger, S. Dawson, and G. Valencia, Phys. Rev. Lett. {\bf 67}(1991)2256;
J.Bagger et al, Phys.Rev. {\bf D49}(1994)1246;
V. Koulovassilopoulos, R.S. Chivukula, BUHEP-93-30 (hep-ph/9312317).

\item                    
 J.M. Cornwall, D.N. Levin, and G. Tiktopoulos, Phys. Rev. {\bf D10}
(1974)1145; C.E. Vayonakis, Lett. Nuovo {\bf 17}(1976)383; B.W. Lee,
C. Quigg, and H. Thacker, Phys. Rev. {\bf D16}(1977)1519.

\item                     
M.S. Chanowitz and M.K. Gaillard, Nucl. Phys. {\bf 261}(1985)379;\\
G.J. Gounaris, R. K$\ddot{o}$gerler, and H. Neufeld, Phys. Rev. {\bf D34}
(1986)3257.

\item                      
Y.P. Yao and C.P. Yuan, Phys. Rev. {\bf D38}(1988)2237;\\
J. Bagger and C. Schmidt, Phys. Rev.  {\bf D41}(1990)264.

\item                      
H.J. He, Y.P. Kuang, and X. Li, Phys. Rev. Lett. {\bf 69}(1992)2619;\\
"{\it Further Investigation on the Precise Formulation
of the Equivalence Theorem} ",
Tsinghua Univ. Preprint TUIMP-TH-92/51
( to be published in Phys. Rev. {\bf D}(1994) ).

\item                       
H.J. He, Y.P. Kuang, and X. Li, "
{\it Precise Formulation of the Equivalence Theorem beyond the
Standard Model} ", VPI-IHEP-93-16 and TUIMP-TH-94/57 ( in preparation ).

\item                      
T. Appelquist and C. Bernard, Phys. Rev. {\bf D22}(1980)200;\\
J. Gasser and H. Leutwyler, Ann. Phys. {\bf 158}(1984)142.

\item                      
See, for example, W.A. Bardeen, Phys. Rev. {\bf D18}(1978)1969.

\item                      
C. Becchi, A. Rouet, B. Stora, Comm. Math. Phys. {\bf 42}(1975)127; Ann.
of Phys. {\bf 98}(1976)287; T. Kugo and I. Ojima, Suppl. Prog. Theor.
Phys. {\bf 66}(1979)1; L. Baulieu, Phys. Rep. {\bf 129}(1985)1.

\item                       
S. Weinberg, Physica {\bf 96A}(1979)327.

\item                       
J.F. Donoghue and J. Tandean, Phys. Lett. {\bf B301}(1993)372.

\item                        
H.J. He and Y.P. Kuang, CCAST Report, CCAST-93-IR-03, June 1993 (unpublished).

\item                        
P.B. Pal, Phys. Lett. {\bf B231}(1994)229;
W.B. Kilgore, LBL Preprint LBL-34900 (hep-ph/9311379).

\item                        
G.G. Sutherland, Nucl. Phys. {\bf B2}(1967)433;
M. Veltman, Proc. Roy. Soc. {\bf A301}(1967)107.
\end{enumerate}

\end{sf}
\end{document}